%% file: esSpinStrain.tex
\documentclass[prl,aps,preprint,superscriptaddress,showpacs]{revtex4-1}
\usepackage[]{graphicx}
\usepackage[]{hyperref}
\usepackage[]{amsmath}
\usepackage{braket}
\usepackage{color}
\usepackage[subrefformat=parens,labelformat=parens]{subcaption}
\usepackage[table]{xcolor}

\begin{document}

\title{Cooling a Mechanical Resonator with a Nitrogen-Vacancy Center Ensemble Using a Room Temperature Excited State Spin-Strain Interaction}

\author{E. R. MacQuarrie}
\affiliation{Cornell University, Ithaca, NY 14853}
\author{M. Otten}
\affiliation{Cornell University, Ithaca, NY 14853}
\author{S. K. Gray}
\affiliation{Argonne National Laboratory, Lemont, IL 60439}
\author{G. D. Fuchs}
\email{gdf9@cornell.edu}
\affiliation{Cornell University, Ithaca, NY 14853}

\begin{abstract}

We propose a protocol to dissipatively cool a room temperature mechanical resonator using a nitrogen-vacancy (NV) center ensemble. The spin ensemble is coupled to the resonator through its orbitally-averaged excited state, which has a spin-strain interaction that has not been previously characterized. We experimentally demonstrate that the spin-strain coupling in the excited state is $13.5\pm0.5$ times stronger than the ground state spin-strain coupling. We then theoretically show that this interaction combined with a high-density spin ensemble enables the cooling of a mechanical resonator from room temperature to a fraction of its thermal phonon occupancy. 

\end{abstract}

\maketitle

Cooling a mechanical resonator to a sub-thermal phonon occupation can uncover exciting physics normally obscured by thermal excitations or enhance sensing by lowering the thermal noise floor~\cite{albrecht1991,cleland1998,rugar2004}. Taken to the extreme, cooling a mechanical mode to the ground state of its motion enables the exploration of quantum effects at the mesoscopic scale~\cite{cleland2010,chan2011,teufel2011}. This has motivated researchers in the field of optomechanics to invent methods for cooling mechanical resonators through their interactions with light. Such techniques have been used to cool resonators to their ground state from cryogenic starting temperatures~\cite{chan2011,teufel2011} and to near their ground state from room temperature~\cite{metzger2004,gigan2006,arcizet2006,kleckner2006,schliesser2006}.

A well-controlled quantum system coupled to the motion of a resonator can also be used to cool a mechanical mode~\cite{rabl2009,kepesidis2013}. For the nitrogen-vacancy (NV) center in diamond, this coupling can arise through coherent interactions with lattice strain~\cite{MacQuarrie2013,teissier2014,ovartchaiyapong2014,MacQuarrie2015,barfuss2015,MacQuarrie2015b,golter2016,lee2016}. These interactions have stimulated the development of single-crystal diamond mechanical resonators~\cite{mitchell2015,burek2015,khanaliloo2015,meesala2016} and several theoretical proposals for cooling such resonators with a single NV center~\cite{rabl2009,kepesidis2013,ma2016,lau2016}. In principle, replacing the single NV center with a many-NV ensemble can provide a collective enhancement to the strain coupling, which would increase the cooling power of these protocols. In practice, however, ensembles can shorten coherence times and introduce inhomogeneities, which may prevent collective enhancement depending on the proposed mechanism.

In this work, we study the hybrid quantum system composed of an ensemble of NV center spins collectively coupled to a mechanical resonator with the goal of developing a method for cooling the resonator from ambient conditions. Experimentally, we characterize the previously unstudied spin-strain coupling within the room temperature NV center excited state (ES) and find that it is $13.5\pm0.5$ times stronger than the ground state (GS) spin-strain interaction. We then propose a dissipative cooling protocol that does not require long spin coherence and theoretically show that a dense NV center ensemble can cool a high-$Q$ mechanical resonator from room temperature to a fraction of its thermal phonon population. 

To achieve substantial cooling from ambient conditions, we require a strong, room temperature NV center-strain interaction that can be enhanced by an ensemble. We first consider the orbital-strain coupling that exists within the NV center ES at cryogenic temperatures. This $850\pm130$~THz/strain interaction offers a promising route towards single NV center-mechanical resonator hybrid quantum systems~\cite{golter2016,lee2016}, but for ensemble coupling, strain inhomogeneities strongly broaden the orbital transition and prohibit collective enhancement. Moreover, the orbital coherence begins to dephase above $10$~K due to phonon interactions~\cite{fu2009}, limiting applications to cryogenic operation.

A weaker ($21.5\pm1.2$~GHz/strain) spin-strain coupling exists at room temperature within the NV center GS~\cite{ovartchaiyapong2014}. The resonance condition for this interaction is determined by a static magnetic bias field which, unlike strain, can be very uniform across an ensemble. This GS spin-strain interaction thus offers a path towards coupling an ensemble to a mechanical resonator. As the NV center density grows, however, the GS spin coherence will decrease~\cite{acosta2009}, limiting the utility of the collective enhancement. 

Finally, we consider spin-strain interactions in the room temperature ES, which have not been thoroughly investigated but which might provide the desired compatibility with dense ensembles. For temperatures above $\sim150$~K, orbital-averaging from the dynamic Jahn-Teller effect erases the orbital degree of freedom from the NV center ES Hamiltonian, resulting in an effective orbital singlet ES at room temperature~\cite{batalov2009,rogers2009,fu2009,plakhotnik2015}. Previous magnetic spectroscopy has hinted that a transverse spin-strain coupling exists within this room temperature ES and that this coupling is $\mathcal{O}(10)$ times stronger than the GS coupling~\cite{fuchs2008,doherty2013}. Like the GS spin-strain interaction, a static magnetic bias field will determine the resonance condition for this coupling, enabling collective enhancement with an ensemble. Furthermore, the NV center density is not expected to affect the ES coherence time, which is limited by the ES motional narrowing rate~\cite{fuchs2010,plakhotnik2015}. This ES spin-strain interaction thus offers a promising path towards an ensemble-mechanical resonator hybrid quantum system. 

\begin{figure}[ht]
\begin{center}
\begin{tabular}{c}
\includegraphics[width=\linewidth]{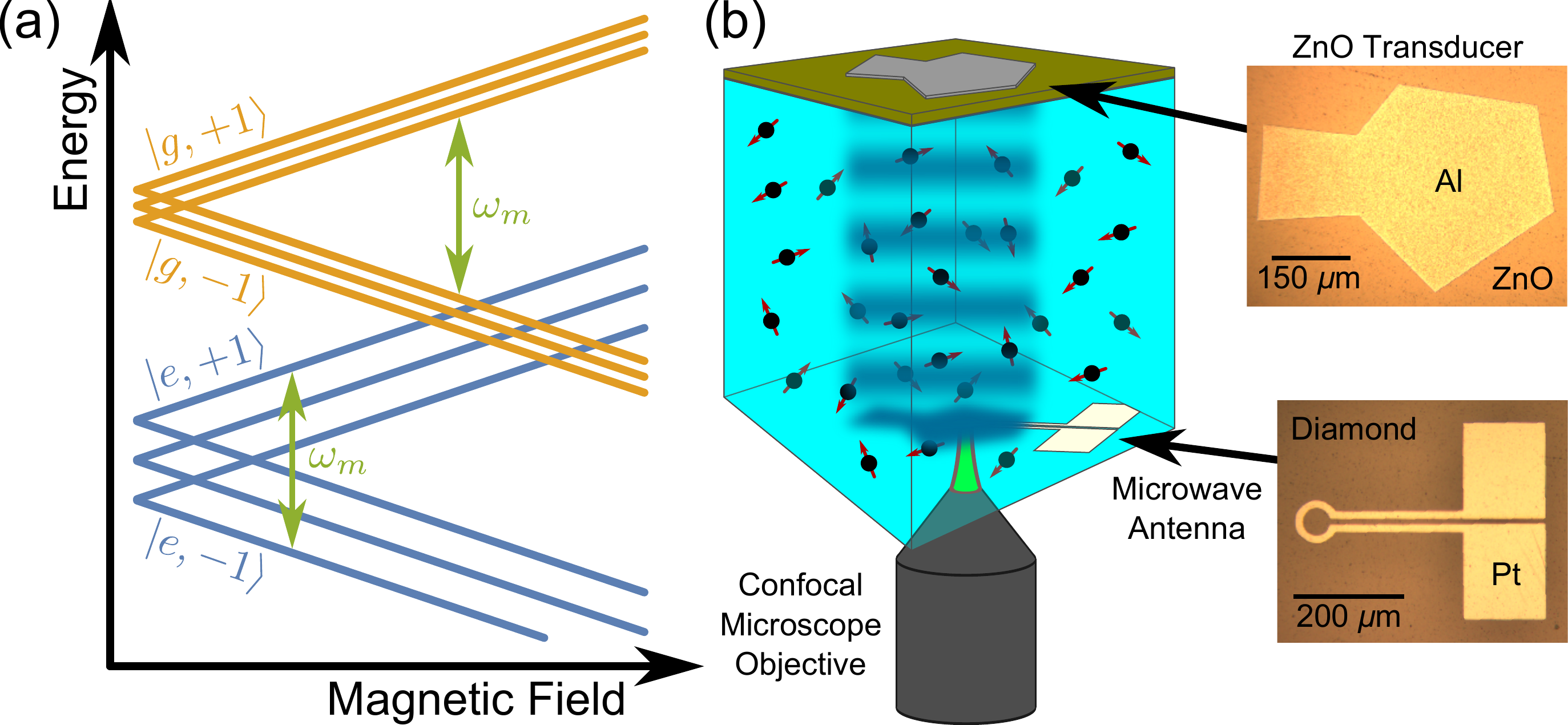} \\
\end{tabular} 
\end{center}
\caption[fig:fig1] {(a) NV center GS and ES energy levels as a function of the magnetic bias field. Energies have been plotted relative to the $m_s=0$ state in each orbital, and a mechanical mode of frequency $\omega_m$ has been drawn connecting the $m_I=+1$ hyperfine sublevels. (b) Schematic of the device used in these measurements.}
\label{fig:fig1}
\end{figure}

To precisely characterize the strength of this ES spin-strain interaction, we start by writing the relevant Hamiltonian in the presence of a magnetic field $\vec{B}$ and strain $\epsilon_{x}$. Both the GS and room temperature ES spin Hamiltonians then take the form ($\hbar=1$)~\cite{Slichter1990,doherty2013}
\begin{equation}
H=D_0 S_z^2+\gamma_{NV}\vec{S}\cdot\vec{B}-d_{\perp}\epsilon_{x}(S_x^2-S_y^2)+A_{\|}S_z I_z
\end{equation}
where $D_0^{e}/2\pi = 1.42$~GHz and $D_{0}^{g}/2\pi=2.87$~GHz are the ES and GS zero-field splittings, $\gamma_{NV}/2\pi=2.8$~MHz/G is the NV center gyromagnetic ratio, $A_{\|}^{e}/2\pi=+40$~MHz~\cite{steiner2010} and $A_{\|}^g/2\pi=-2.166$~MHz are the ES and GS hyperfine couplings to the $^{14}$N nuclear spin, $\vec{S}$ ($\vec{I}$) is the electronic (nuclear) spin-$1$ Pauli vector, and the $z$-axis runs along the NV center symmetry axis. Perpendicular strain $\epsilon_x$ couples the $\Ket{+1}$ and $\Ket{-1}$ spin states with a strength $d_{\perp}^e$ in the ES and $d_{\perp}^g/2\pi=21.5\pm1.2$~GHz/strain in the GS~\cite{ovartchaiyapong2014}. As shown in Fig.~\ref{fig:fig1}a, this interaction allows direct control of the magnetically-forbidden $\Ket{+1}\leftrightarrow\Ket{-1}$ spin transition through resonant strain. 

The combination of a large hyperfine coupling in the ES and a short ES lifetime broadens the spectral features of the ES spin-strain interaction. Measuring such a spectrum then requires large magnetic field sweeps ($\Delta B_z\sim 150$~G), which in turn require a mechanical driving field with a high carrier frequency ($\omega_m/2\pi\gtrsim420$~MHz). To this end, we fabricate a high-overtone bulk acoustic resonator (HBAR) capable of generating large amplitude strain at gigahertz-scale frequencies. Our resonator consists of a ZnO piezoelectric film sandwiched between two electrodes patterned on one face of a $\langle100\rangle$ single-crystal diamond substrate. Applying a high-frequency voltage to this transducer launches acoustic waves into the diamond, which serves as a Fabry-P\'{e}rot cavity to generate a comb of standing wave resonances. The resonator used in this work was driven at a $\omega_m/2\pi=529$~MHz mechanical mode that has a quality factor of $Q=1790\pm20$. A magnetic antenna fabricated on the opposite diamond face provides high-frequency magnetic fields for magnetic spin control. The final device is pictured in Fig.~\ref{fig:fig1}b. 

The CVD-grown diamond used in these measurements contained NV centers at a density of $\sim 4\times 10^{14}$~cm$^{-3}$ as purchased. Our measurements thus address an ensemble of $\sim70$ NV centers oriented with their symmetry axis parallel to a static magnetic bias field $B_z$. NV centers of different orientations are spectrally isolated and contribute only a constant background to the measurements. 

\begin{figure}[ht]
\begin{center}
\begin{tabular}{c}
\includegraphics[width=\linewidth]{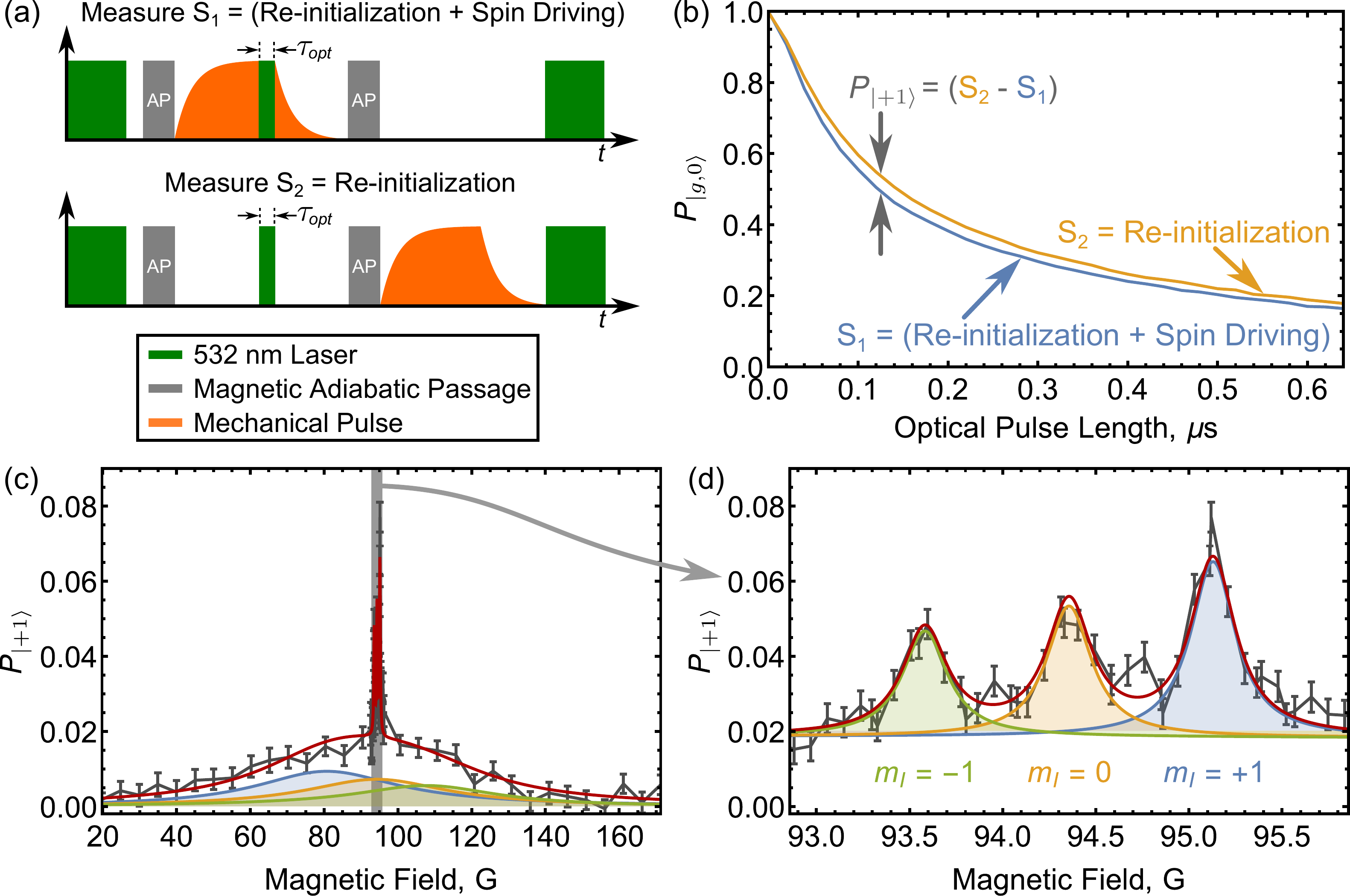} \\
\end{tabular} 
\end{center}
\caption[fig:fig2] {(a) Pulse sequences used to measure ES spin driving. (b) Population in $\Ket{g,0}$ at the end of the pulse sequences in (a) plotted against $\tau_{opt}$. (c) Spectrum of the mechanically driven spin population. (d) Zoomed in view of the GS spin transitions in (c). }
\label{fig:fig2}
\end{figure}

To measure mechanical spin driving within the ES, we execute the pulse sequences shown in Fig.~\ref{fig:fig2}a as a function of $B_z$. In the first pulse sequence, a $532$~nm laser initializes the NV center spin into the GS level $\Ket{g,(m_s=)0}$. A magnetic adiabatic passage (AP) then moves the spin population to $\Ket{g,-1}$. At this point, we turn on the mechanical driving field for $3$~$\mu$s. Just before the end of the mechanical pulse, we apply a $\tau_{opt}=125$~ns optical pulse with a $532$~nm laser. This excites the electron spin to $\Ket{e,-1}$ and allows the spin to interact with the mechanical driving field in the ES. If the driving field is resonant with the $\Ket{e,+1}\leftrightarrow\Ket{e,-1}$ splitting, population will be driven into $\Ket{e,+1}$. The spin then follows either a spin-conserving relaxation down to $\Ket{g,\pm1}$ or a relaxation to the singlet state $\Ket{S_1}$ through an intersystem crossing. The former preserves the spin state information, while relaxing to $\Ket{S_1}$ re-initializes the spin, erases the stored signal, and reduces the overall spin contrast of the measurement. After allowing the spin to relax, we apply a second magnetic AP to return the spin population in $\Ket{g,-1}$ to $\Ket{g,0}$ and measure the $\Ket{g,0}$ spin population via fluorescence read out. We define this signal as $S_1$ and plot it as a function of $\tau_{opt}$ in Fig.~\ref{fig:fig2}b. 

In the second pulse sequence, the mechanical pulse occurs between the second AP and fluorescence read out. Applying the mechanical pulse with the spin in $\Ket{g,0}$ maintains the same power load on the device but does not drive spin population. This sequence measures $S_2$, the re-initialization of the NV center from the $\tau_{opt}$ optical pulse (Fig.~\ref{fig:fig2}b). Subtracting $S_2-S_1$ gives the probability of the spin being in $\Ket{+1}$ at the end of the first sequence. A third sequence with a single AP and a fourth with two APs (both with $\tau_{opt}=0$) normalize the spin contrast at each $B_z$. 

Fig.~\ref{fig:fig2}c shows the results of such a measurement. We fit the data to the sum of six Lorentzians as described in the SI~\cite{SI}. When the mechanical driving field is resonant with the $\Ket{g,+1}\leftrightarrow\Ket{g,-1}$ splitting, population is driven into $\Ket{+1}$ by the GS spin-strain interaction for the duration of the mechanical pulse~\cite{MacQuarrie2013}. Fig.~\ref{fig:fig2}d highlights this GS driving with a zoomed in view of the spectrum about the GS resonances. The reversed sign of $A_{\|}^e$ relative to $A_{\|}^g$ is consistent with \textit{ab initio} calculations~\cite{gali2008} and was confirmed by measurements that were conditional on the nuclear spin state~\cite{SI}. 

\begin{figure*}[ht]
\begin{center}
\begin{tabular}{c}
\includegraphics[width=\linewidth]{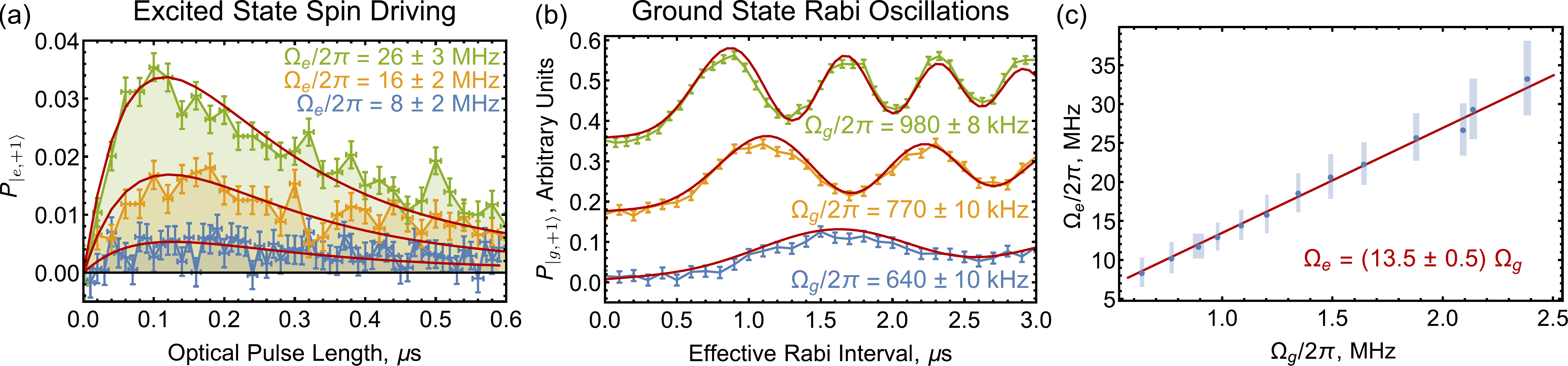} \\
\end{tabular} 
\end{center}
\caption[fig:fig3] {(a) Population in $\Ket{e,+1}$ plotted as a function of $\tau_{opt}$. The red lines are least squares fits to a seven-level master equation model of the measurement. (b) Mechanically driven Rabi oscillations within the GS orbital. (c) The ES mechanical driving field plotted against the GS mechanical driving field.}
\label{fig:fig3}
\end{figure*}

To quantify the strength of the ES spin-phonon interaction, we spectrally isolate the ES transition by fixing $B_z=80$~G and execute a modified version of the pulse sequence described above. Here, we use $\sim 20$~ns magnetic $\pi$-pulses to address the $\Ket{g,0}\leftrightarrow\Ket{g,-1}$ transition and measure both $S_1$ and $S_2$ as functions of $\tau_{opt}$ and of the power applied to the HBAR. As Fig.~\ref{fig:fig3}a shows, taking $S_2-S_1$ reveals the competition between mechanical spin driving into $\Ket{e,+1}$ and re-initialization of the spin state via optical pumping. The overlaid lines are least squares fits of the data to a seven-level master equation model. From the fits, we extract the value of the ES mechanical driving field $\Omega_e$. The SI includes a detailed description of this model, which was designed to account for inhomogeneities within the NV center ensemble and for the polarization of the nuclear spin sublevels, among other effects~\cite{SI}. By mechanically driving Rabi oscillations within the NV center GS (Fig.~\ref{fig:fig3}b), we measure the GS mechanical driving field $\Omega_g$ for each power applied to the HBAR~\cite{SI}. Plotting $\Omega_e$ against $\Omega_g$ (Fig.~\ref{fig:fig3}c) shows that the transverse spin-strain coupling in the ES is $13.5\pm0.5$ times stronger than the GS coupling, or $d_{\perp}^e/2\pi=290\pm20$~GHz/strain. 

\begin{figure}[ht]
\begin{center}
\begin{tabular}{c}
\includegraphics[width=\linewidth]{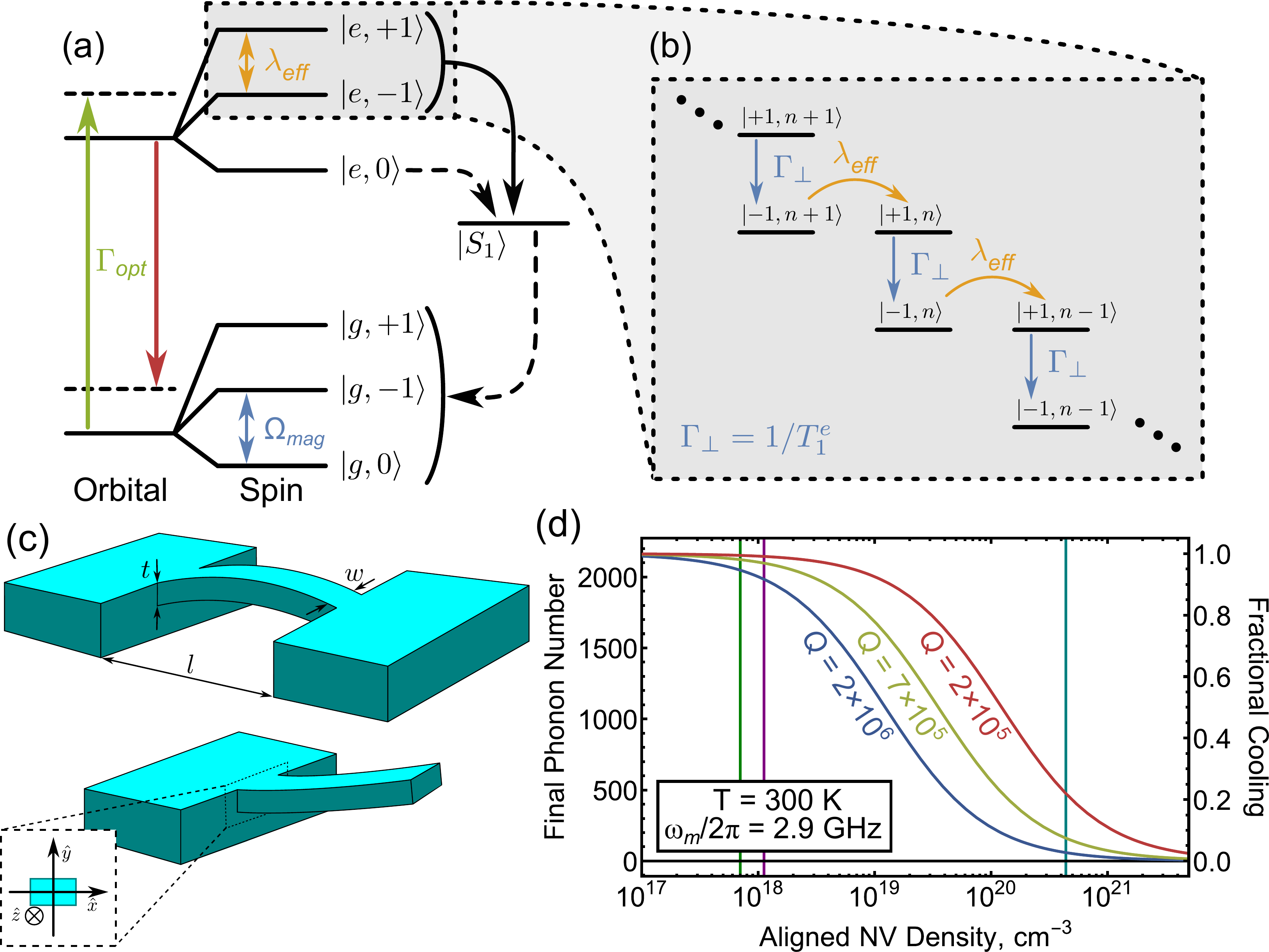} \\
\end{tabular} 
\end{center}
\caption[fig:fig4] {(a) The seven NV center orbital and spin states at room temperature. Fast (slow) transitions are indicated by solid (dashed) one-way arrows. Coherent couplings are indicated by two-way arrows. (b) Simplified depiction of the proposed cooling protocol. (c) Schematics of a doubly-clamped beam (top) and a cantilever (bottom). (d) Final phonon number achieved by the cooling protocol as a function of $\rho$. Vertical lines demarcate densities that have been realized in single-crystal diamonds ($7\times10^{17}$~cm$^{-3}$~\cite{acosta2009}, $1.1\times10^{18}$~cm$^{-3}$~\cite{wee2007}) and nanodiamonds~($4.4\times10^{20}$~cm$^{-3}$~\cite{silkin2011}). }
\label{fig:fig4}
\end{figure}

With $d_{\perp}^e$ quantified, we will now present a protocol for cooling a mechanical resonator with an NV center spin ensemble. In our proposed protocol, a $532$~nm laser continuously pumps the phonon sidebands of the ensemble's optical transition, and a gigahertz frequency magnetic field continuously drives the $\Ket{g,0}\leftrightarrow\Ket{g,-1}$ spin transition. This generates a steady state population surplus in $\Ket{e,-1}$ as compared to $\Ket{e,+1}$, enabling the net absorption of phonons by the ensemble. Spontaneous relaxation and subsequent optical pumping continually re-initialize the system, allowing the phonon absorption cycle to continue. Fig.~\ref{fig:fig4}a summarizes this process. 

Under the Tavis-Cummings model, an ensemble of $N$ two-level systems coupled to a phonon mode with strength $\lambda$ can be described by a single harmonic oscillator coupled with an effective strength $\lambda\sqrt{N}$~\cite{tavis1968,bullough1987,garraway2011}. In order to cast our proposed protocol into this framework, we approximate the seven-level NV center landscape with the two-level toy model shown in Fig.~\ref{fig:fig4}b. 

To justify this simplification, we compared numerical simulations of the cooling from a small number of seven-level NV centers to that predicted analytically by the Tavis-Cummings toy model. Due to the exponential growth of the Hilbert space, simulations were performed on the Titan supercomputer at Oak Ridge National Laboratory, with the most intensive simulations taking $\sim 10^4$ core-hours. As described in the SI, we determined that the two-level model provides an upper bound on the steady state phonon number $n_f$, suggesting that the proposed protocol cools a resonator more efficiently than our toy model predicts~\cite{otten2015, SI}. This motivates further theoretical study into extensions of the Tavis-Cummings model. 

Within the toy model, the ensemble-resonator dynamics are described by the master equation ($\hbar=1$)
\begin{equation}
\dot{\rho}=-i[H,\rho]+\mathcal{L}_{\Gamma}\rho+\mathcal{L}_{\gamma}\rho
\label{eq:mastercool}
\end{equation}
where $\mathcal{L}_{\Gamma}$ describes the incoherent NV center processes, $\mathcal{L}_{\gamma}$ describes the resonator rethermalization, and $H$ describes the coherent coupling between the spin ensemble and the resonator. For resonant coupling, the quantized Hamiltonian in the Tavis-Cummings form is~\cite{tavis1968,garraway2011}
\begin{equation}
H=\omega_m(a^{\dagger}a+J_{+}J_{-}) + \lambda_{eff}(J_+ +J_-)(a^{\dagger}+a)
\end{equation}
where $a^{\dagger}$ ($a$) is the creation (annihilation) operator for the mechanical mode, $J_{\pm}$ are the ladder operators for the spin state, and $\lambda_{eff}$ is the effective spin-phonon coupling strength. The spin relaxation term in Eq.~\ref{eq:mastercool} takes the form $\mathcal{L}_{\Gamma}\rho=\left(2T_{1}^e\right)^{-1}\mathcal{D}[J_-]\rho+\left(2T_2^e\right)^{-1}(J_z \rho J_z-\rho)$ where $\mathcal{D}[J_-]\rho=(2 J_-\rho J_+-J_+J_-\rho-\rho J_+J_-)$ is the Lindblad superoperator, $T_2^e=6.0$~ns is the ES coherence time~\cite{fuchs2012}, and $T_{1}^e=6.89$~ns is the ES lifetime of $\Ket{e,+1}$~\cite{robledo2011}. The resonator rethermalization is described by $\mathcal{L}_{\gamma}\rho=\frac{\gamma}{2}(n_{th}+1)\mathcal{D}[a]\rho+\frac{\gamma}{2}n_{th}\mathcal{D}[a^{\dagger}]\rho$ where $\gamma=\omega_m/Q$ is the mechanical dissipation rate and $n_{th}\sim k_B T/\hbar \omega_m$ is the thermal phonon occupancy of the resonator mode. 

We make the approximation that at some initial time, the mechanical mode is in a coherent state $\Ket{a^{\dagger}a}=\Ket{n_{th}}$. Solving for the steady state matrix of second order moments then provides an analytic expression for $n_{f}$ ~\cite{wilsonrae2008,SI}, allowing us to study the cooling performance as a function of the device parameters.

Elasticity theory provides a means of predicting those device parameters for doubly-clamped beams and cantilevers. For resonators of length $l$, thickness $t$, and width $w$, we compute the strain due to the zero-point motion of the resonator $\epsilon_0\left(y,z\right)$ with coordinates as defined in Fig.~\ref{fig:fig4}c. Assuming a uniform distribution of properly aligned NV centers at a density $\rho$,~\cite{schuster2010,bennett2013}
\begin{equation}
\lambda_{eff}= d_{\perp}^e\sqrt{\alpha \rho w}\sqrt{\int_0^l \int_{-t/2}^{t/2} \epsilon_0^2(y,z) dy dz}
\label{eq:gEff}
\end{equation}
where $\alpha$ is the steady state population difference between $\Ket{e,-1}$ and $\Ket{e,+1}$. To calculate $\alpha$, we solve for the $7\times 7$ density matrix describing the steady state of the NV center ensemble in the absence of the mechanical resonator and obtain $\alpha=0.017$ for reasonable control fields~\cite{SI}. Evaluating Eq.~\ref{eq:gEff}, we find that $\lambda_{eff}$ is independent of $w$ and scales as \makebox{$\lambda_{eff}= G_0\sqrt{t}/l$} where $G_0=d_{\perp}^e\sqrt{\hbar\kappa_0\alpha\rho/E}$, $\kappa_0^b=120$~GHz$\cdot\mu$m for beams, $\kappa_0^c=19$~GHz$\cdot\mu$m for cantilevers, and $E=1200$~GPa is the Young's modulus of diamond. The frequency of the fundamental mode scales as \makebox{$\omega_m=\kappa_0 t/l^2$}. For transform-limited linewidths, higher order mechanical modes are spectrally isolated from the NV center spin dynamics for the resonators considered here~\cite{bennett2013,ovartchaiyapong2014,SI}. At room temperature, phonon-phonon interactions limit the $Q$ of an ideal diamond mechanical resonator. For modes in the resolved-sideband regime ($\omega_m/2\pi>1/T_2^e\sim170$~MHz), the maximum $Q=2\times 10^6$ is independent of $\omega_m$ at a fixed bath temperature~\cite{tabrizian2009}.
 
Assuming fully polarized nuclear spins, the fractional cooling $n_{f}/n_{th}$ is insensitive to the physical dimensions of any resonator satisfying $\omega_m/2\pi>1/T_2^e$ because the size of the ensemble scales with the size of the resonator. For illustrative purposes, we will examine a $\omega_m/2\pi=2.9$~GHz resonator. This fixes $B_z$ near the ES level anti-crossing where dynamic nuclear polarization is most effective~\cite{fuchs2008,jacques2009}. Taking $Q=2\times 10^6$, the optimum dimensions then become $(l,t)=(3.3,1.6)$~$\mu$m for a beam and $(l,t)=(0.51,0.26)$~$\mu$m for a cantilever. Wee, \textit{et al} reported measurements of an NV center ensemble with $\rho=1.1\times10^{18}$~cm$^{-3}$ in single-crystal diamond~\cite{wee2007}. For this density, we find that the proposed protocol cools a room temperature resonator to $n_{f}\sim 0.92 n_{th}$. 

We can further reduce $n_f$ by increasing $\rho$. The magnetic field noise from paramagnetic impurities grows with $\rho$, which will degrade the GS coherence time. However, for large magnetic driving fields, this cooling protocol does not require a lengthy GS coherence time~\cite{SI}. The only coherence time that effects the protocol is $T_2^e$, which is not expected to change with the defect density. This means that enormously high NV center densities could in principle be used to cool a resonator with the ES spin-strain interaction. To study how increasing $\rho$ affects the protocol, we plot $n_{f}$ against $\rho$ in Fig.~\ref{fig:fig4}d for several different $Q$-values. For reference, we have included lines marking values of $\rho$ that have been realized in single-crystal diamonds~\cite{acosta2009,wee2007} and in nanodiamonds~\cite{silkin2011}. The limiting density of NV centers in a single-crystal diamond nanostructure is currently unknown. Furthermore, while high defect densities have been shown to degrade the $Q$ of $\omega_m/2\pi\sim10$~kHz frequency resonators~\cite{tao2014}, it remains to be seen how the gigahertz frequency resonators of interest here will be affected by the incorporation of a dense defect ensemble. These questions motivate future experimental work.

Finally, we note that another approach to lowering $n_f$ could be to simultaneously implement an optomechanical cooling protocol that uses radiation pressure to reduce the phonon occupancy of the resonator. By supplementing our proposed protocol with such techniques, cooling a mechanical resonator from room temperature to its quantum ground state may become possible. 

In conclusion, we have proposed a dissipative resonator cooling protocol that utilizes an ensemble of NV center spins to realize a collective enhancement in the spin-phonon coupling. After demonstrating that the spin-strain coupling in the room temperature ES is $13.5\pm0.5$ times stronger than the GS spin-strain coupling, we analyzed the performance of the proposed protocol. For very dense NV center ensembles, the ES cooling protocol can cool a room temperature resonator to a fraction of its thermal phonon occupancy. These results demonstrate the exciting opportunities collectively coupled ensembles offer NV center-mechanical resonator hybrid quantum systems and shed further light on the orbitally-averaged room temperature ES of the NV center. 

\vspace*{2mm}

\input acknowledgment.tex 

\section{Supplementary Information}

\subsection{Spectrum Fitting}

The spectrum pictured in Fig.~\ref{fig:fig2} of the main text was fit to the function
\begin{equation}
\begin{split}
P_{\Ket{+1}}&=c_e\left(\frac{a_+[B_z]}{\frac{1}{4}\Gamma_e^2+(B_z-B_0+A_{\|}^e/\gamma_{NV})^2} + \frac{a_0[B_z]}{\frac{1}{4}\Gamma_e^2+(B_z-B_0)^2} + \frac{a_-[B_z]}{\frac{1}{4}\Gamma_e^2+(B_z-B_0-A_{\|}^e/\gamma_{NV})^2}\right) \\
&+ c_g\left(\frac{a_+[B_z]}{\frac{1}{4}\Gamma_g^2+(B_z-B_0+A_{\|}^g/\gamma_{NV})^2} + \frac{a_0[B_z]}{\frac{1}{4}\Gamma_g^2+(B_z-B_0)^2} + \frac{a_-[B_z]}{\frac{1}{4}\Gamma_g^2+(B_z-B_0-A_{\|}^g/\gamma_{NV})^2}\right)
\end{split}
\label{eq:Spect}
\end{equation}
where $c_{e}$ and $c_{g}$ are constant amplitudes that quantify the driven spin contrast, $a_i[B_z]$ is a field-dependent scaling factor that accounts for the dynamic nuclear polarization of the hyperfine sublevels~\cite{jacques2009}, $\Gamma_{e}$ ($\Gamma_{g}$) is the FWHM of the excited (ground) state resonances, $B_0$ is the resonant field for the $m_I=0$ hyperfine sublevel, and the other parameters are as defined in the main text. Of these variables, $c_i$, $\Gamma_i$, and $B_0$ are free parameters in our fitting procedure. 

\begin{figure}[ht]
\begin{center}
\begin{tabular}{c}
\includegraphics[width=\linewidth]{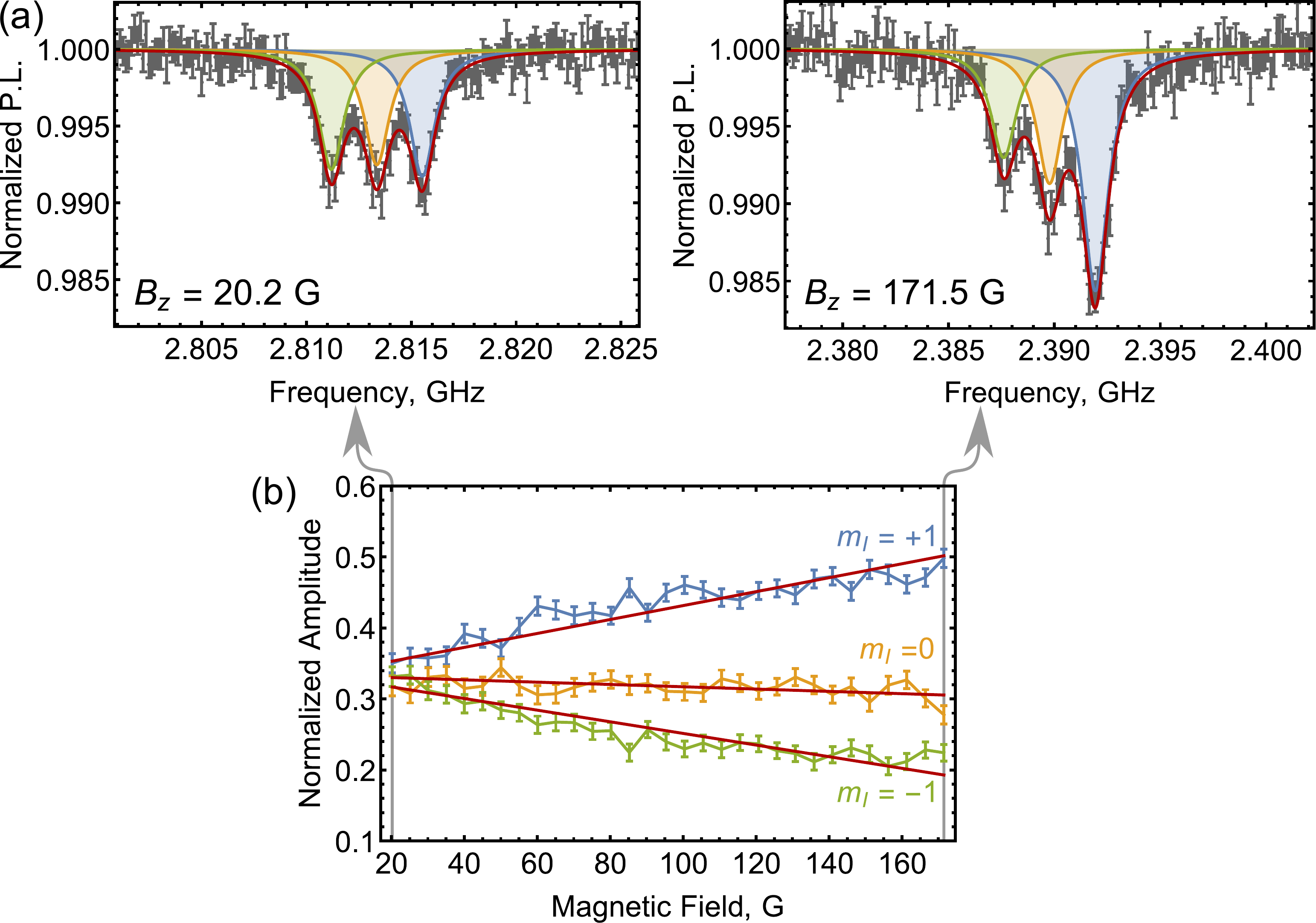} \\
\end{tabular} 
\end{center}
\caption[fig:SpectFig] {(a) Normalized photoluminescence plotted as a function of the magnetic driving field carrier frequency. (b) Normalized amplitude of each hyperfine sublevel as a function of $B_z$. }
\label{fig:SpectFig}
\end{figure}

We calibrate $a_i[B_z]$ by performing hyperfine-resolved magnetic electron spin resonance (ESR) measurements within the NV center ground state (GS) as a function of magnetic field. We fit the resulting curves to the function 
\begin{equation}
\begin{split}
P&= C\left(\frac{A_+}{\frac{1}{4}\Gamma_g^2+(B_z-B_0+A_{\|}^g)^2} + \frac{A_0}{\frac{1}{4}\Gamma_g^2+(B_z-B_0)^2} + \frac{A_-}{\frac{1}{4}\Gamma_g^2+(B_z-B_0-A_{\|}^g)^2}\right)+P_0
\end{split}
\end{equation}
where $P$ is the measured photoluminescence, $C$ accounts for the driven spin contrast, $P_0$ is the background photoluminescence, and we fix $\sum A_{i} = 1$. Fig.~\ref{fig:SpectFig}a shows the resulting ESR curves at $B_z=20.2$~G and $B_z=171$~G. We have used the values of $P_0$ returned from the fits to normalize the photoluminescence. As expected, the nuclear polarization increases in the direction of the excited state (ES) level anti-crossing at $B_z^{LAC}=507$~G.

Fig.~\ref{fig:SpectFig}b shows each $A_i$ plotted as a function of $B_z$. We have fit each curve to a straight line with a fixed $y$-intercept of $\frac{1}{3}$ to obtain the linear scaling functions $a_i[B_z]$ in Eq.~\ref{eq:Spect}. The sum of these scaling functions satisfies $\sum a_i[B_z]=1$. 

\subsection{Sign of $A_{\|}$}

\begin{figure}[ht]
\begin{center}
\begin{tabular}{c}
\includegraphics[width=\linewidth]{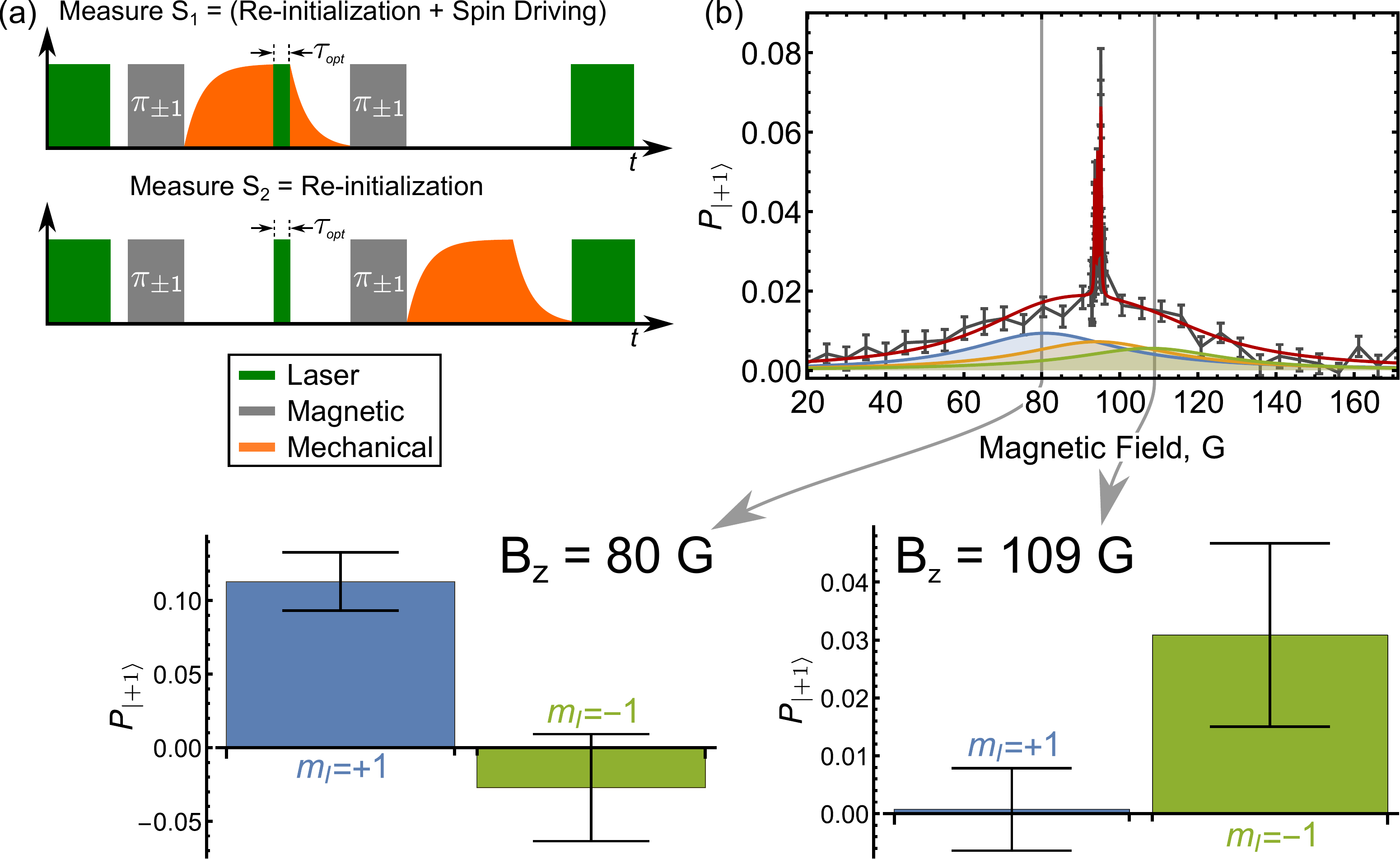} \\
\end{tabular} 
\end{center}
\caption[fig:hfSign] {(a) Pulse sequence used to verify the sign of $A_{\|}^e$. (b) Mechanically driven spin contrast as a function of $B_z$. The bottom two plots show measurements taken at the high-field and low-field hyperfine resonances that were conditional on the $^{14}$N nuclear spin state. }
\label{fig:hfSign}
\end{figure}

To confirm the reversal in the sign of the hyperfine coupling $A_{\|}$ between the GS and ES orbitals, we mechanically drive spin population in the ES conditional on the spin state of the $^{14}$N nucleus. These measurements follow the pulse sequence depicted in Fig.~\ref{fig:hfSign}a. This modified pulse sequence replaces the hard $\pi$-pulses used to quantify $d_{\perp}^e
$ in the main text with weak $\pi$-pulses conditional on the nuclear spin state. 

As shown in Fig.~\ref{fig:hfSign}b, we perform this measurement at the high-field and low-field ES hyperfine resonances. We observe mechanically-driven spin population in the $m_I=+1$ manifold at the $B_z=80$~G resonance and in the $m_{I}=-1$ manifold at the $B_z=109$~G resonance. The resonance condition for spin driving is $\omega_m=2\gamma_{NV}B_z+2A_{\|}^e m_{I}$, giving $A_{\|}^e=\left(\frac{1}{2}\omega_m-\gamma_{NV}B_z\right)/m_{I}$. Using the parameter values given in the main text, this gives $A_{\|}^e/2\pi=+40$~MHz. 

\subsection{Seven-Level Master Equation Model}

\begin{figure}[ht]
\begin{center}
\begin{tabular}{c}
\includegraphics[width=\linewidth]{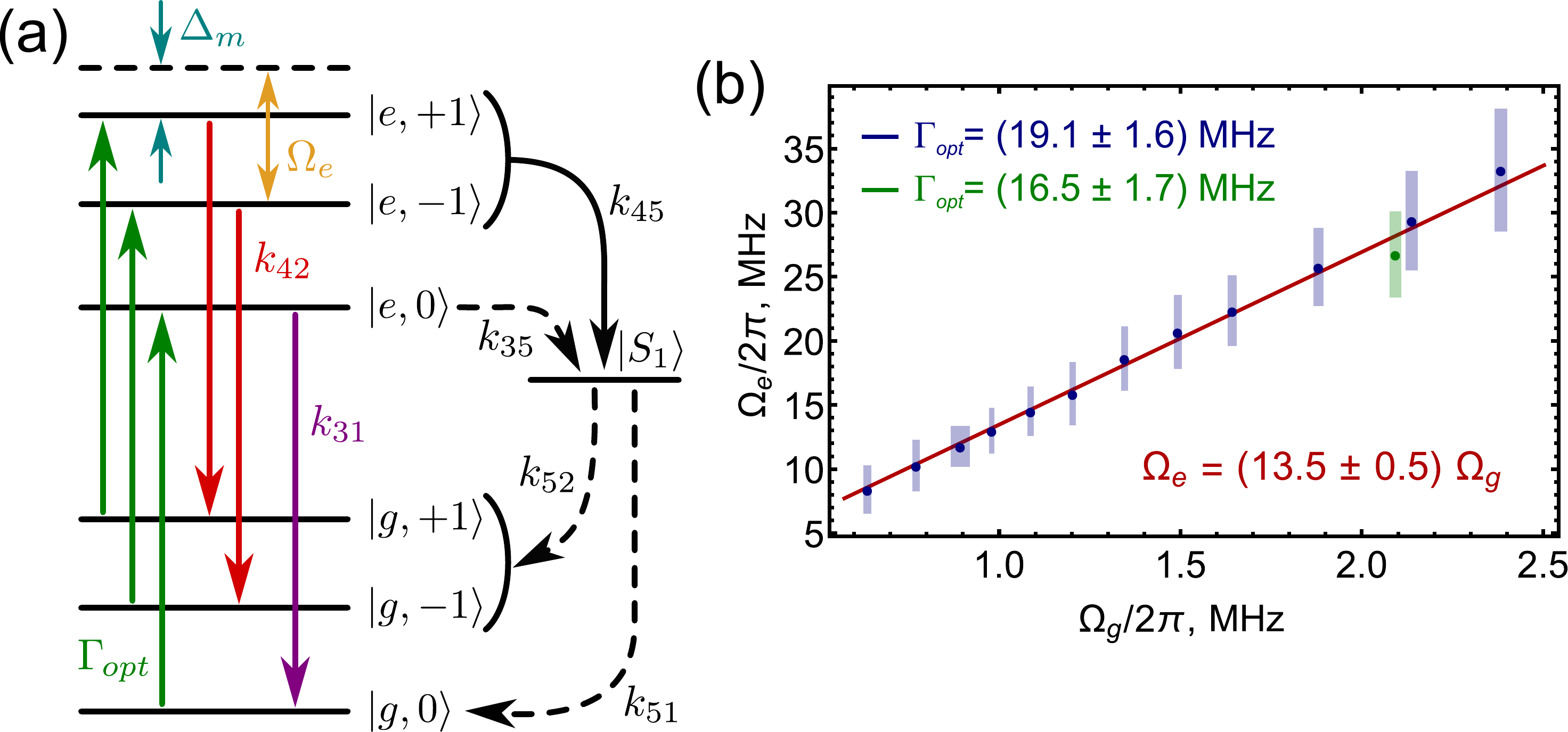} \\
\end{tabular} 
\end{center}
\caption[fig:nvLevels]{(a) States and transitions included in our seven-level master equation model. (b) ES mechanical driving field plotted as a function of the GS mechanical driving field with the data labeled by the optical pumping rate $\Gamma_{opt}$ during each measurement. }
\label{fig:nvLevels}
\end{figure}

The master equation used to model our ES spin driving measurements is derived in the room temperature NV center basis defined by the states $\{g_{+1},g_0,g_{-1},e_{+1},e_0,e_{-1},S_1\}$ where a subscript denotes the $m_s$ value. The $7\times 7$ density matrix evolves according to ($\hbar=1$)
\begin{equation}
\dot{\rho}=-i[H,\rho]+\mathcal{L}_{\Gamma}\rho.
\label{eq:evolve}
\end{equation}
In the rotating frame, the Hamiltonian is given by 
\begin{equation}
\begin{split}
H=\frac{\Omega_e}{2}\left(\Ket{e_{+1}}\Bra{e_{-1}}+\Ket{e_{-1}}\Bra{e_{+1}}\right)+\Delta_m\Ket{e_{+1}}\Bra{e_{+1}}
\end{split}
\end{equation}
where $\Delta_m$ is the mechanical detuning. The incoherent NV center processes are described by
\begin{equation}
\begin{split}
\mathcal{L}_{\Gamma}\rho=&\Gamma_{opt}\sum\limits_{i=\pm1,0}L_{g_i,e_i}+k_{42}\sum\limits_{i=\pm1}L_{e_i,g_i} +k_{45}\sum\limits_{i=\pm1}L_{e_i,S_1}+k_{52}\sum\limits_{i=\pm1}L_{S_1,g_i} \\
&+k_{31}L_{e_0,g_0}+k_{35}L_{e_0,S_1}+k_{51}L_{S_1,g_0}+\frac{1}{T_2^e}\sum\limits_{i=\pm1,0}L_{e_i,e_i}
\end{split}
\end{equation}
where we define
\begin{equation}
L_{i,f}\rho=\Ket{f}\bra{i}\rho\Ket{i}\Bra{f}-\frac{1}{2}\left(\Ket{i}\Bra{i}\rho+\rho\Ket{i}\Bra{i}\right).
\end{equation}
Here, $\Gamma_{opt}$ is the optical pumping rate of our $532$~nm laser, $T_2^e=6.0\pm0.8$~ns is the ES coherence time~\cite{fuchs2012}, and the relaxation rates $k_{ij}$ are listed in Table~\ref{tab:rates}. Fig.~\ref{fig:nvLevels}a summarizes this landscape. 

\begin{table}
\begin{center}
\rowcolors{1}{cyan}{white}
\begin{tabular}{c|c|c}
\textbf{Parameter}	& \textbf{Value}	& 	\textbf{Relaxation From:} \\
$k_{42}$ 	& $65.3\pm1.6$~MHz 	& ES to GS\\
$k_{31}$ 	& $64.9\pm1.5$~MHz 	& ES to GS\\
$k_{45}$ 	& $79.8\pm1.6$~MHz 	& ES to $\Ket{S_1}$\\
$k_{35}$ 	& $10.6\pm1.5$~MHz 	& ES to $\Ket{S_1}$\\
$k_{52}$ 	& $2.61\pm0.06$~MHz 	& $\Ket{S_1}$ to GS\\
$k_{51}$ 	& $3.00\pm0.06$~MHz 	& $\Ket{S_1}$ to GS
\end{tabular}
\end{center}
\caption[tab:rates]{Relaxation rates in our seven-level master equation model~\cite{robledo2011}.}
\label{tab:rates}
\end{table}

Because optical initialization does not generate a pure state, we first simulate the optical pumping process to obtain an initialized density matrix. To do so, we start with a thermal state $\rho_{NV}=\frac{1}{3}\left(\sum\limits_{i=\pm1,0}\Ket{g_i}\Bra{g_i}\right)$ and apply Eq.~\ref{eq:evolve} with $\Omega_e,\Delta_m=0$ and $\Gamma_{opt}\neq0$ for $10$~$\mu$s. We take the resulting density matrix and apply Eq.~\ref{eq:evolve} for $5$~$\mu$s with $\Omega_e,\Delta_m,\Gamma_{opt}=0$ to simulate the relaxation to $\Ket{g}$. A simulated $\pi$-pulse then swaps $\rho_{22}$ and $\rho_{33}$, providing the appropriate starting density matrix $\rho_0$ for a given $\Gamma_{opt}$. From $\rho_0$, we also extract the minimum and maximum spin contrast ($s_{min}=\Bra{g_{0}}\rho_{0}\Ket{g_0}$ and $s_{max}=\Bra{g_{-1}}\rho_{0}\Ket{g_{-1}}$), which allow us to properly normalize our simulations. 

We begin by modeling the measurement of $S_2$, the spin re-initialization. To do so, we apply Eq.~\ref{eq:evolve} to $\rho_0$ with $\Omega_e,\Delta_m=0$ and $\Gamma_{opt}\neq0$ for a length of time $\tau_{opt}$. Allowing the spin to relax as before gives us the measured density matrix $\rho_2$. We normalize  $\Bra{g_{-1}}\rho_{2}\Ket{g_{-1}}$ using $s_{min}$ and $s_{max}$, and repeat this simulation as a function of $\tau_{opt}$ to obtain a simulated $S_2$ curve. 

To account for spatial inhomogeneities in the optical power within the spin ensemble, we perform a weighted average of this simulation over the point spread function (PSF) of our microscope. We approximate the PSF by the function 
\begin{equation}
\Gamma_{opt}(z)=\Gamma_{0}\bigg\lbrace\frac{\sin\lbrace\kappa[z_0](z-z_0)\rbrace}{\kappa[z_0](z-z_0)}\bigg\rbrace^2
\label{eq:psf}
\end{equation}
where $\Gamma_0$ is the peak optical pumping rate, $\kappa[z_0]$ defines the depth-dependent PSF width~\cite{MacQuarrie2013}, $z$ is the distance below the diamond surface, and $z_0=7.9\pm0.9$~$\mu$m is the focus depth of the PSF. An ensemble measurement is then given by
\begin{equation}
S_2^{ens}(\tau_{opt})=\frac{\int_{0}^{\infty}S_2[\tau_{opt},\Gamma_{opt}(z)]dz}{\int_{0}^{\infty}\Gamma_{opt}(z)dz}.
\end{equation}
We discretize this integral to make it computationally tractable and perform a least squares fit of $S_2^{ens}(\tau_{opt})$ to the measured data. $\Gamma_{opt}$ is the only free parameter in the fitting procedure. 

With the exception of the datum indicated in Fig.~\ref{fig:nvLevels}b, all of the measurements were taken at the same optical power. We simultaneously fit each of these $S_2$ curves and measure $\Gamma_{opt}=19.1\pm1.6$~MHz. For the measurement at a different optical power, we measure $\Gamma_{opt}=16.5\pm1.7$~MHz. 

To extract $\Omega_e$, we then fix $\Gamma_{opt}$ and repeat this procedure with $\Omega_e\neq0$ to simulate the $S_2-S_1$ measurement pictured in Fig.~3a of the main text. To account for inhomogeneities in the mechanical driving field, we must also include the spatial profile of the strain standing wave inside the weighted average. This function takes the form $\Omega_e=\Omega_0 |\sin\left[2\pi z/\lambda\right]|$ where $\lambda=31\pm4$~$\mu$m is the wavelength of the strain wave. Defining the results of such a simulation as $P_{\Ket{+1}}\left(\Omega_e,\Delta_m\right)$, we account for the hyperfine sublevels by computing the sum 
\begin{equation}
P_{\Ket{+1}}\left(\Omega_e\right)=a_{+1} P_{\Ket{+1}}\left(\Omega_e,0\right)+a_{0}P_{\Ket{+1}}\left(\Omega_e,A_{\|}\right)+a_{-1}P_{\Ket{+1}}\left(\Omega_e,2A_{\|}\right)
\label{eq:finalModel}
\end{equation}
where the normalized amplitudes ($\sum a_i=1$) account for nuclear spin polarization and have been measured separately via magnetic ESR. A least squares fit of Eq.~\ref{eq:finalModel} to the data then provides $\Omega_e$. Here, $\Omega_e$ is the only free parameter in the fitting procedure. 

When fitting the relation between $\Omega_e$ and $\Omega_g$ (Fig.~\ref{fig:fig3}c of the main text), we fix the $y$-intercept of the linear fitting function to be $0$. 

\subsection{GS Mechanically Driven Rabi Oscillations}

\begin{figure}[ht]
\begin{center}
\begin{tabular}{c}
\includegraphics[width=\linewidth]{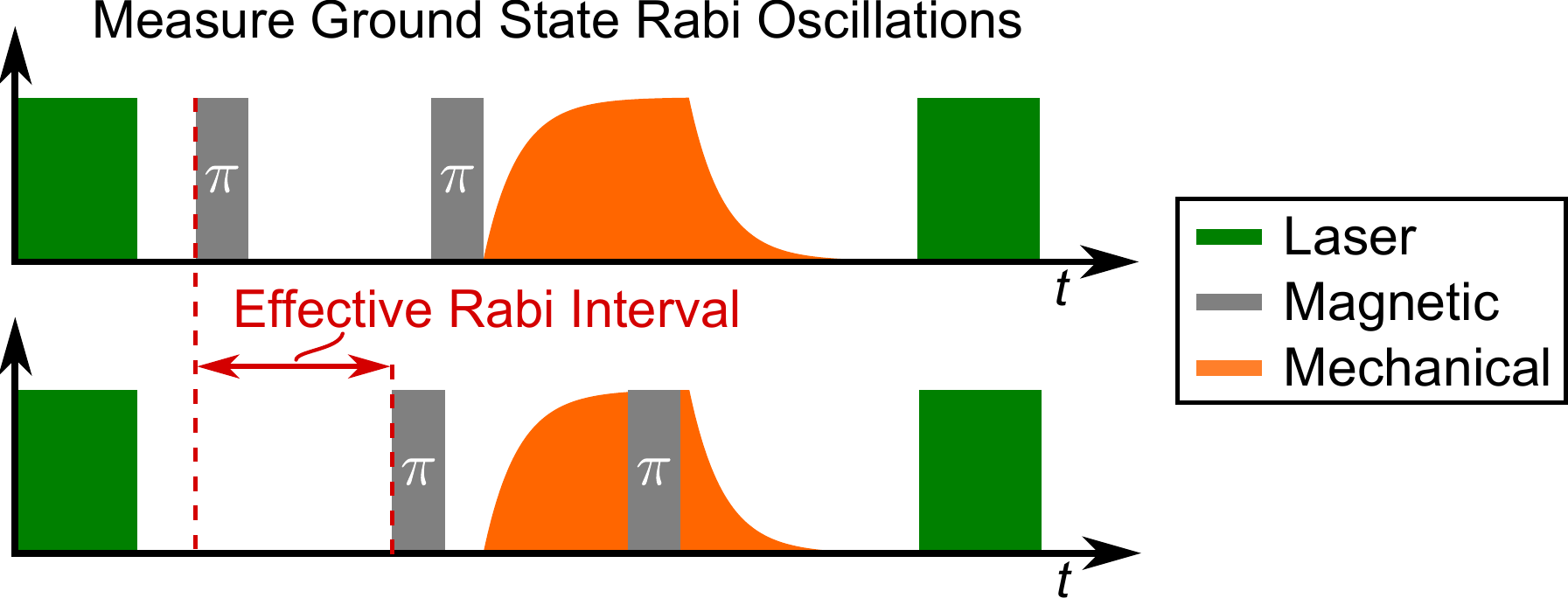} \\
\end{tabular} 
\end{center}
\caption[fig:gsRabi]{Pulse sequence used to measure mechanically driven GS Rabi oscillations. }
\label{fig:gsRabi}
\end{figure}

The GS mechanically driven Rabi oscillations used to quantify $\Omega_g$ were measured using the pulse sequence shown in Fig.~\ref{fig:gsRabi}. As described in detail in Ref.~\cite{MacQuarrie2015}, varying the pulse length of our mechanical driving field introduces bandwidth-related artifacts to a Rabi measurement. Instead, we fix the length of the mechanical pulse and vary the interaction time by sweeping a pair of magnetic $\pi$-pulses through the mechanical pulse. This yields the data seen in Fig.~\ref{fig:fig3}b of the main text where the ``Effective Rabi Interval'' label on the $x$-axis corresponds to the delay of the $\pi$-pulse pair. 

For a single NV center, a GS Rabi measurement is described by the function
\begin{equation}
P_{\Ket{+1}}(t,\Omega_g)=\frac{1}{2} \big\lbrace 1 - e^{-t/T_{Rabi}} \cos[\Omega_g \tau(t)]\big\rbrace
\label{eq:gsRabi}
\end{equation}
where $\tau(t)=\int_{0}^{t}\left(1-e^{-t'/\tau_Q}\right)dt'=\left[\left(e^{-t/\tau_{Q}}-1\right) \tau_{Q}+t\right]$ accounts for the ring-up of the mechanical resonator and $\tau_Q=2Q/\omega_0$. 

For an ensemble measurement, we account for spatial inhomogeneities by taking the weighted average of Eq.~\ref{eq:gsRabi}: 
\begin{equation}
P_{\Ket{+1}}^{ens}=\frac{\int_{0}^{\infty}P_{\Ket{+1}}(t,\Omega_g|\sin\left[2\pi z/\lambda\right]|)\Gamma_{opt}(z)dz}{\int_{0}^{\infty}\Gamma_{opt}(z)dz}
\label{eq:gsRabiEns}
\end{equation}
where $\Gamma_{opt}(z)$ is as defined in Eq.~\ref{eq:psf} above. We discretize this integral and fit each GS Rabi curve to Eq.~\ref{eq:gsRabiEns}, fixing $\tau_Q$ to be the same across the fits. 

\section{Numerical Validation of Two-Level Model}

As the main text explains (around Eq.~3 of the main text), we make a simplification from the 
full seven-level system to an effective two-level system. This allows us to make use of the
Tavis-Cummings model to do calculations of large ensembles ($10^5$ NVs) by replacing the 
coupling, $\lambda$, by an effective coupling $\sqrt{n_{nv}}\lambda$, where $n_{nv}$ is the number
of NVs. Without Tavis-Cummings, the calculations would be completely intractable, since
the Hilbert space size grows exponentially with $n_{nv}$. 

This simplification has two potential sources
of error: the reduction from the seven-level system to the two-level system for one NV and
the scaling from one NV to many NVs. To understand the error bounds, we solve the
seven-level model explicitly, solving for the steady state using the
full Hilbert space, similar to the method used in~\cite{otten2015}. 
To efficiently solve for the steady state, ($\mathbf{A}x = 0$), we represent the 
master equation in superoperator space and explicitly construct the $\mathbf{A}$ matrix as
\begin{equation}\label{superoperator}
\mathbf{A}x = (-i (\mathbf{I} \otimes H - H \otimes \mathbf{I}) + \tilde{L})x,
\end{equation}
where $x=vec(\rho)$, the vectorization of $\rho$, constructed by stacking the 
columns of $\rho$ into a single column vector,
$H$ is the Hamiltonian of the system, $\mathbf{I}$ is the identity matrix in the 
total Hilbert space, and $\tilde{L}$ is the collection of Lindblad superoperators represented
in superoperator space. For example, using the single Lindblad superoperator 
$\gamma D[a]\rho = \gamma (2 a \rho a^\dagger - a^\dagger a \rho - \rho a^\dagger a)$, gives
\begin{equation}\label{l_tilde}
\tilde{L} = \gamma (2 a \otimes a - \mathbf{I} \otimes a^\dagger a - a^\dagger a \otimes \mathbf{I}).
\end{equation}
The superoperator form of both $H$ and the Lindblad terms are derived from the 
identity $\mathbf{A}\mathbf{X}\mathbf{B} = (\mathbf{B^T} \otimes \mathbf{A}) vec(\mathbf{X})$,
where $\mathbf{A}, \mathbf{B},$ and $\mathbf{X}$ are all matrices.
$H$ in Eq.~\ref{superoperator} can generally
represent any system, but we used the seven-level Hamiltonian of Eq.~\ref{eq:evolve} 
(or multiple instances of this, in the case of more than one NV). 

$\mathbf{A}$ is an extremely large matrix, but it is also extremely sparse. For our system,
$\mathbf{A}$ has less than 10 non-zeros per row, but, for the seven-level Hamiltonian, has 
matrix dimensions $N^2 \times N^2,$ where $N=7^{n_{nv}}n_{ph}, n_{nv} =$ 
number of NVs and $n_{ph} = $
the largest phonon state accessible in the simulation. We utilize the software package 
PETSc~\cite{petsc-user-ref,petsc-efficient} to perform these large, but very sparse, 
calculations. $\mathbf{A}$ is stored in compressed sparse row format, ensuring we 
do the minimal amount of calculations and use the minimum amount of storage. 
 $\mathbf{A}$ is a complex, nonsymmetric matrix, restricting us to use
GMRES~\cite{Saad1986} as our parallel iterative solver, which has slow convergence, especially
with increasing system size. Explicitly constructing $\mathbf{A}$ allows us to use 
efficient preconditioners, such as the additive Schwarz method, to accelerate
the convergence. We solve for the steady state rather than doing explicit time dynamics
because of the wide separation of time scales in our model. The NV dynamics are very 
fast, while the cooling is much slower. For an explicit time stepping approach, 
millions of time steps are necessary to get to the steady state solution, whereas the 
steady state results typically converge in less than 5000 iterations.

To understand the error from the model reduction, we first focus on simulations using just 
a single NV. To see significant cooling in manageable computational time, we increase the spin-strain coupling by a factor of 100 and reduce the resonator frequency to $\omega_m/2\pi=475$~MHz, causing observable cooling but ensuring that the resonator is still only
a small perturbation upon the NV center dynamics. We also restrict ourselves to small 
$N_{th}$ values (equivalently, small temperatures) so that the Hilbert space size needed
to approximate the infinite phonon bath is small and the computation remains tractable.
 
\begin{figure}[ht]
\begin{center}
\includegraphics[width=\linewidth]{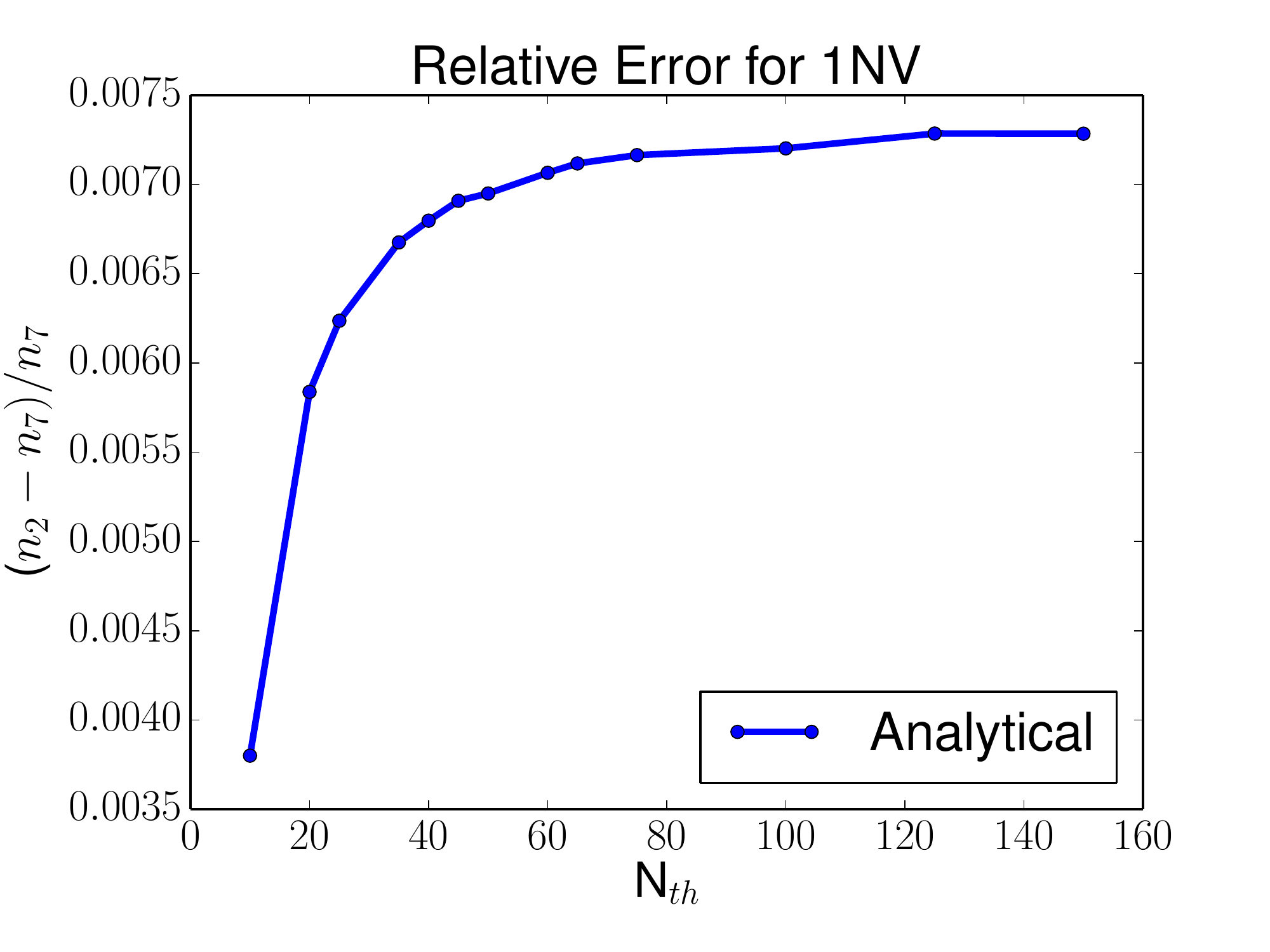} \\
\end{center}
\caption[fig:1nv_comparison]{Error in the analytic two-level simplification compared to the
  full seven-level system for one NV.}
\label{fig:1nv_comparison}
\end{figure}

Figure \ref{fig:1nv_comparison} shows the relative error of the two-level system 
with respect to the seven-level system with only one NV. 
As $N_{th}$ increases, the error increases and then
levels off at an asymptotic value of less than 0.0075, less than 1\% error. At room temperature,
where $N_{th}$ is on the order of 10,000 this simplification would only give an error 
of only 75 phonons in the final phonon number, showing that the two-level reduction
is justified, at least for a single NV. Furthermore, it is important to note that the simplified two-level model underestimates the cooling as compared to the seven-level
model. The simplified two-level model thus serves as an upper bound on the final phonon number $n_f$. 

\begin{figure}[ht]
  \centering
  \begin{subfigure}{0.5\textwidth}
    \centering
    \includegraphics[width=\columnwidth]{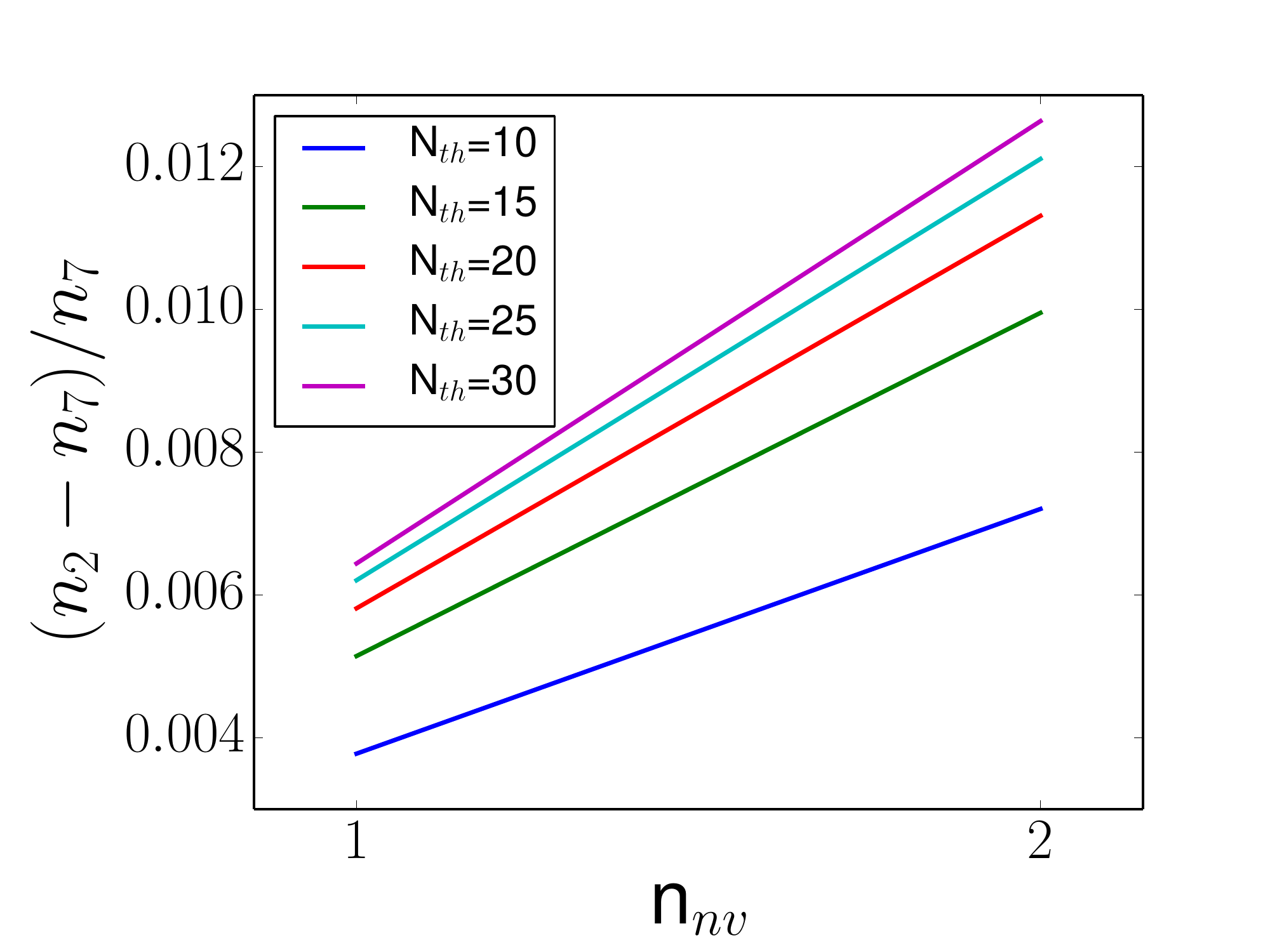}
    \caption{
      Relative error of the two state system with respect to the seven state system
      for different numbers of NVs and different $N_{th}$ values.
    }\label{many_nv(a)}
  \end{subfigure}%
  ~
  \begin{subfigure}{0.5\textwidth}
    \centering
    \includegraphics[width=\columnwidth]{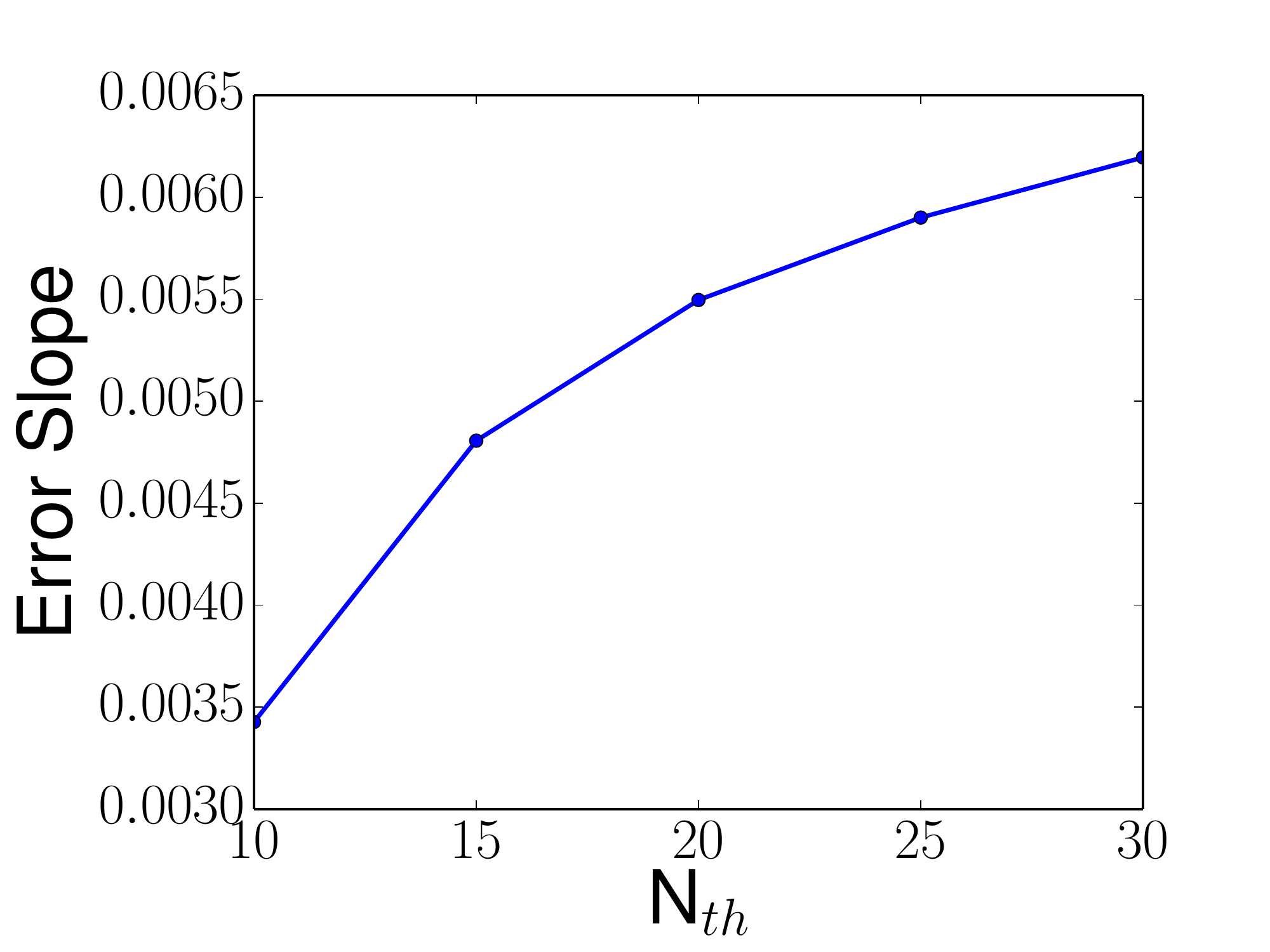}
    \caption{
      Slope of fitted lines to the different $N_{th}$ values in figure~\ref{many_nv(a)}.
    }\label{many_nv(b)}
  \end{subfigure}%
  \caption{
    Results for one and two NVs.
  }
\end{figure}

It is also important to understand how the error scales when the number of NVs is increased. This is a much more computationally
challenging task, since each additional seven-level NV increases the total Hilbert space
size by a factor of seven. It is only feasible to use two or three NVs, in addition to
the mechanical resonator. As such, we did several calculations with one and two NVs, and
a few, small, $N_{th}$ values, as show in figure~\ref{many_nv(a)}. The error gets worse
going from one NV to two NVs, but the slope of this change is different for the 
different $N_{th}$ values. In fact, the slope seems to converge with increasing
$N_{th}$, as shown in figure~\ref{many_nv(b)}. While it is hard to make any 
conclusion about the error for $10^5$ NVs, we can at least see that the increase in error
is such that the two-level simplification is still an upper bound to the cooling,
though a slightly worse one. We also did calculations with three NVs where possible, 
and verified that the linear behavior extends to
at least three NVs. Explicitly including enough NVs to see the many NV behavior
is computationally intractable, and motivates further theoretical study, such as
investigations into extensions to the Tavis-Cummings model for systems with
more than two states. Nevertheless, these results imply that $n_f$ predicted by the simplified two-level model serves as an upper bound, and the protocol may cool better than the toy model suggests. 

\subsection{Steady State Phonon Occupancy}

Within the two-level toy model used to analyze the proposed cooling protocol, the ensemble-resonator dynamics are governed by
\begin{equation}
\dot{\rho}=-i[H,\rho]+\mathcal{L}_{\Gamma}\rho+\mathcal{L}_{\gamma}\rho
\end{equation}
where $H$, $\mathcal{L}_{\Gamma}\rho$, and $\mathcal{L}_{\gamma}\rho$ are defined in the main text. Using these expressions and the ladder operator commutation relations, we derive the system of ordinary differential equations that governs the time evolution of the second order moments~\cite{wilsonrae2008}. This system is given by
\begin{equation}
\begin{split}
&\frac{d}{dt} \langle a^{\dagger}a \rangle  = \left(-i\lambda_{eff}\right)\left(\langle J_+a^{\dagger} \rangle -\langle J_-a\rangle - \langle J_+a \rangle+\langle J_-a^{\dagger} \rangle\right) + \gamma n_{th}-\gamma \langle a^{\dagger}a \rangle,\\
&\frac{d}{dt} \langle J_+J_- \rangle = \left(-i\lambda_{eff}\right)\left(\langle J_+a^{\dagger} \rangle -\langle J_-a\rangle - \langle J_-a^{\dagger} \rangle+\langle J_+a \rangle\right) - \Gamma_{\perp} \langle J_+J_- \rangle,\\
&\frac{d}{dt} \langle J_+a^{\dagger} \rangle = \left(i\lambda_{eff}\right)\left(1+\langle a^{\dagger} a^{\dagger}\rangle + \langle J_+J_+\rangle+\langle J_+J_-\rangle+\langle a^{\dagger}a\rangle\right) - \left(\frac{1}{2}\Gamma_{\perp}+\frac{1}{2}\gamma-2 i \omega_m+\Gamma_{\|}\right)\langle J_+ a^{\dagger}\rangle, \\
&\frac{d}{dt} \langle J_-a^{\dagger} \rangle = \left(-i\lambda_{eff}\right)\left(\langle a^{\dagger}a\rangle+\langle a^{\dagger}a^{\dagger} \rangle - \langle J_+J_-\rangle - \langle J_-J_-\rangle\right)-\left(\frac{1}{2}\Gamma_{\perp}+\frac{1}{2}\gamma+\Gamma_{\|}\right)\langle J_- a^{\dagger}\rangle, \\
&\frac{d}{dt} \langle J_+ a \rangle = \left(i\lambda_{eff}\right)\left(\langle a^{\dagger}a\rangle+\langle aa\rangle - \langle J_+J_-\rangle - \langle J_+J_+\rangle\right)-\left(\frac{1}{2}\Gamma_{\perp}+\frac{1}{2}\gamma+\Gamma_{\|}\right)\langle J_+ a\rangle,\\
&\frac{d}{dt} \langle J_- a \rangle = \left(-i\lambda_{eff}\right)\left(1+\langle a a\rangle + \langle J_-J_-\rangle+\langle J_+J_-\rangle+\langle a^{\dagger}a\rangle\right) - \left(\frac{1}{2}\Gamma_{\perp}+\frac{1}{2}\gamma+2 i \omega_m+\Gamma_{\|}\right)\langle J_- a\rangle, \\
&\frac{d}{dt} \langle J_-J_-\rangle = \left(-2i\lambda_{eff}\right)\left(\langle J_-a^{\dagger}\rangle+\langle J_- a\rangle\right) - \left(\Gamma_{\perp}+2i\omega_m+\frac{1}{2}\Gamma_{\|}\right)\langle J_-J_-\rangle,\\
&\frac{d}{dt} \langle J_+J_+\rangle = \left(2i\lambda_{eff}\right)\left(\langle J_+a^{\dagger}\rangle+\langle J_+ a\rangle\right) - \left(\Gamma_{\perp}-2i\omega_m+\frac{1}{2}\Gamma_{\|}\right)\langle J_+J_+\rangle,\\
&\frac{d}{dt} \langle a^{\dagger}a^{\dagger}\rangle = \left(2i\lambda_{eff}\right)\left(\langle J_+a^{\dagger}\rangle+\langle J_- a^{\dagger}\rangle\right) - \left(\gamma-2i\omega_m\right)\langle a^{\dagger}a^{\dagger}\rangle,\\
&\text{and}\\
&\frac{d}{dt} \langle a a\rangle = \left(-2i\lambda_{eff}\right)\left(\langle J_+a\rangle+\langle J_- a\rangle\right) - \left(\gamma+2i\omega_m\right)\langle aa\rangle.
\end{split}
\end{equation}
Solving this system of equations gives an unwieldy expression for the steady state phonon occupancy $n_{f}$. We numerically solve this expression to study the performance of the proposed cooling protocols. 

\subsection{Elasticity Theory}

To analyze how strain couples to NV centers within a resonator, we start by assuming that the NV centers are aligned with the direction of beam deflection such that the strain in an oscillating beam is entirely perpendicular to the NV center symmetry axis. We then use elasticity theory to derive the scaling laws quoted in the main text~\cite{bennett2013,ovartchaiyapong2014}. 

The wave equation for both doubly-clamped beams and cantilevers is
\begin{equation}
\rho_d A \frac{\partial^2}{\partial t^2}\phi(t,z)=-E I\frac{\partial^4}{\partial z^4}\phi(t,z)
\end{equation}
where $\phi(t,z)$ is the transverse displacement in the $y$-direction, $\hat{z}$ is along the beam as indicated in Fig.~\ref{fig:fig4}c of the main text, $A=wt$ is the cross-sectional area of the resonator, $E=1200$~GPa is the Young's modulus of diamond, $\rho_d=3.515$~g/cm$^{3}$ is the mass density of diamond, and $I=wt^3/12$ is the resonator's moment of inertia. Solutions are of the form $\phi(z,t)=u(z)e^{-i\omega t}$ where
\begin{equation}
u_n(z)=a_n\left(\cos k_n z-\cosh k_n z\right) - b_n\left(\sin k_n z - \sinh k_n z\right).
\end{equation}
The allowed $k$-vectors satisfy $\cos\left(k_n z\right) \cosh \left(k_n z\right)=-1$ for a cantilever and $\cos\left(k_n z\right) \cosh \left(k_n z\right)=1$ for a beam. For a cantilever, the wave vector and amplitudes of the fundamental mode satisfy $k_0^cl=1.875$ and $a_0/b_0=1.3622$, respectively. For a beam, this becomes $k_0^b l = 4.73$ and $a_0/b_0=1.0178$. 

We normalize $u_n\left(z\right)$ by setting the free energy of the beam equal to the zero point energy of the mode:
\begin{equation}
W=\frac{1}{2}EI\int_0^L\left(\frac{\partial^2 u_n}{\partial z^2}\right)^2dz = \frac{1}{2}\hbar\omega_n
\end{equation}
where the eigenfrequencies of the resonator are given by $\omega_n=k_n^2\sqrt{EI/\rho_d A}$. For the fundamental mode, this becomes $\omega_0=\kappa_0 t/l^2$ where $\kappa_0=(k_0 l)^2\sqrt{E/12\rho_d}$, giving $\kappa_0^c=19$~GHz$\cdot\mu$m for cantilevers and $\kappa_0^b=120$~GHz$\cdot\mu$m for beams as quoted in the main text. 

The spin-strain coupling for a single NV center located at $\left(y,z\right)$ is given by $\lambda_{s}=d_{\perp}\epsilon_0\left(y,z\right)$ where $\epsilon_0\left(y,z\right)=-y\frac{\partial^2}{\partial z^2}u_n(z)$ is the strain from the zero point motion of the resonator mode. Here, the $y$-axis is zeroed at the neutral axis of the resonator. To compute the ensemble-resonator coupling, we assume a uniform distribution of properly aligned NV centers within the resonator and take the weighted average of the individual couplings in quadrature. This gives 
\begin{equation}
\lambda_{eff}=d_{\perp}^e\sqrt{\alpha \rho l t w}\sqrt{\int_0^l \int_{-t/2}^{t/2} \epsilon_0^2(y,z) dy dz\bigg/\int_0^l \int_{-t/2}^{t/2} dy dz},
\end{equation}
which simplifies to Eq.~\ref{eq:gEff} of the main text. Evaluating the integrals gives \makebox{$\lambda_{eff}= G_0\sqrt{t}/l$} where $G_0=d_{\perp}\sqrt{\hbar\kappa_0\alpha\rho/E}$ as quoted in the main text. 

\subsection{Higher Order Mechanical Modes}

The frequencies of higher order mechanical modes scale with the resonator dimensions as $\omega_n=\kappa_n t/l^2$. For a beam, $\kappa_n^b=120, 330, 628, ...$~GHz$\cdot\mu$m, while for a cantilever, $\kappa_n^c=19, 120, 330, ...$~GHz$\cdot\mu$m. 

We limit our analysis to $\omega_0/2\pi=2.9$~GHz resonators. The next order mode of such a resonator will be $\omega_1^b/2\pi=8.0$~GHz for a beam and $\omega_1^c/2\pi=18.3$~GHz for a cantilever. Assuming a transform-limited ES linewidth of $\sim1/T_2^e=170$~MHz, the resulting spectral isolation is more than enough to isolate the NV center ensemble from the higher order mechanical modes. 

\subsection{Required Control Fields for Cooling}

\begin{figure}[ht]
\begin{center}
\begin{tabular}{c}
\includegraphics[width=\linewidth]{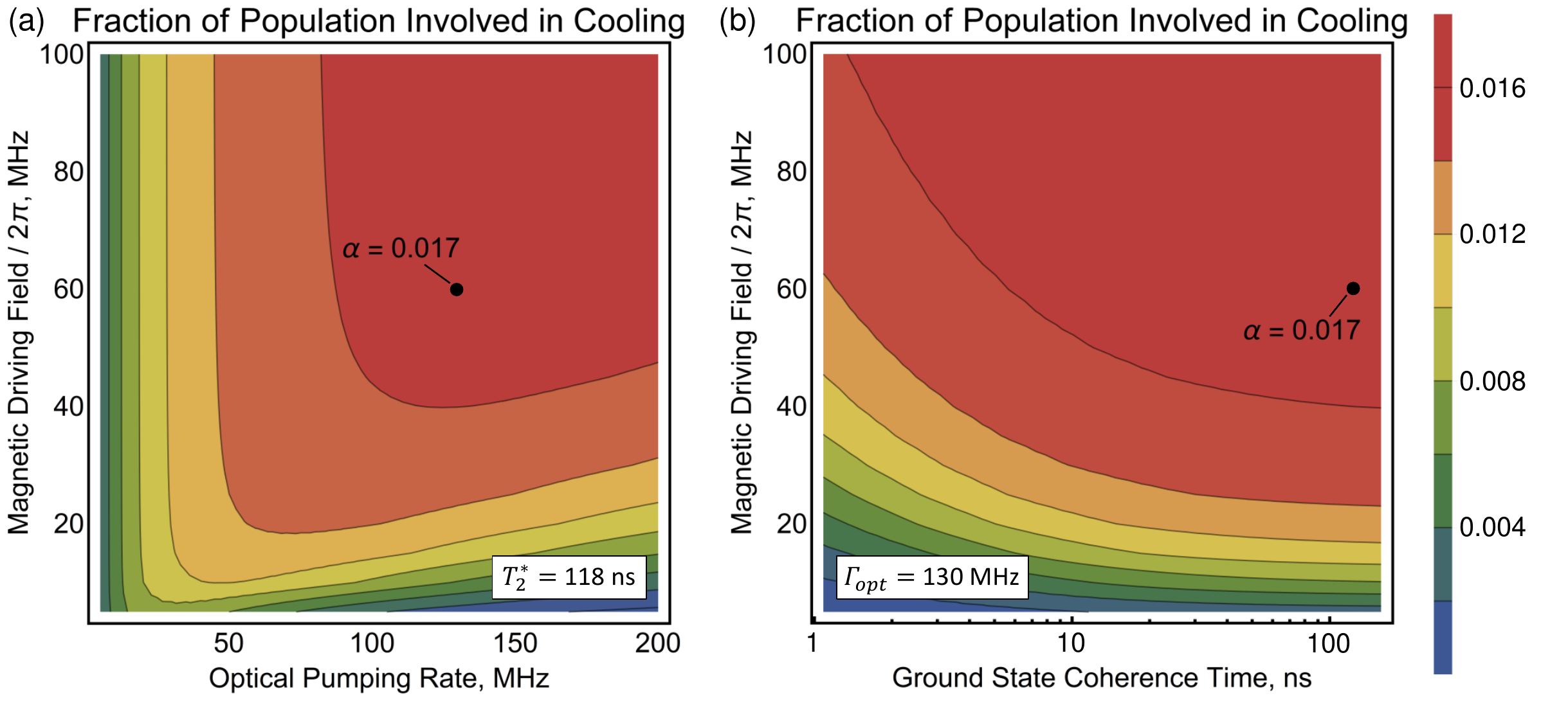} \\
\end{tabular} 
\end{center}
\caption[fig:contra]{The fraction of the ensemble population involved in the cooling plotted as a function of $\Omega_{mag}$ and (a) $\Gamma_{opt}$ or (b) the ground state coherence time.}
\label{fig:contra}
\end{figure}

The proposed cooling protocol requires a static magnetic bias field $B_z$ to bring the spin-strain interaction into resonance, a continuous gigahertz-frequency magnetic field $\Omega_{mag}$ to address the $\Ket{g,0}\leftrightarrow\Ket{g,-1}$ transition, and continuous optical illumination $\Gamma_{opt}$ to re-initialize the system. 

For the $\omega_0/2\pi=2.9$~GHz resonators considered in this work, $B_z=\left(\omega_m/2-A_{\|}\right)/\gamma_{NV}=500$~G. The magnitudes of $\Omega_{mag}$ and $\Gamma_{opt}$ determine $\alpha$, the fraction of the ensemble population involved in the cooling. A large $\Gamma_{opt}$ is desired to saturate the steady state population in the NV center ES. A large $\Omega_{mag}$ is also required to maximize the spin population driven into $\Ket{g,-1}$. As shown in Fig.~\ref{fig:contra}a, $\alpha$ saturates for large control fields at $\alpha\sim 0.017$. For our analysis of the cooling protocol, we use $\Omega_{mag}/2\pi=60$~MHz and $\Gamma_{opt}=130$~MHz. Such a large $\Omega_{mag}$ has been previously demonstrated in ground state spin control experiments~\cite{fuchs2009}. As demonstrated by Fig.~~\ref{fig:contra}b, for $\Omega_{mag}/2\pi=60$~MHz, $\alpha$ remains robust for ground state coherence times $\gtrsim 6$~ns. 

\bibliography{bibDiMEMS}

\end{document}

%% file: acknowledgment.tex
%
Research support was provided by the Office of Naval Research (ONR) (Grant N000141410812). Device fabrication was performed in part at the Cornell NanoScale Science and Technology Facility, a member of the National Nanotechnology Coordinated Infrastructure, which is supported by the National Science Foundation (Grant ECCS-15420819), and at the Cornell Center for Materials Research Shared Facilities which are supported through the NSF MRSEC program (DMR-1120296). Numerical simulations were performed in part at the Center for Nanoscale Materials, a U.S. Department of Energy Office of Science User Facility under Contract No. DE-AC02-06CH11357. This research used resources of the Oak Ridge Leadership Computing Facility at the Oak Ridge National Laboratory, which is supported by the Office of Science of the U.S. Department of Energy under Contract No. DE-AC05-00OR22725.